\documentclass[aps,reprint,prd,twocolumn,floatfix,]{revtex4}%

\usepackage{amsfonts}
\usepackage{amsmath}
\usepackage{amssymb}
\usepackage{graphicx}
\usepackage[ansinew]{inputenc}
\usepackage{pdfpages}
\usepackage{epsfig}
\usepackage{lipsum}
\usepackage{footnote}
\usepackage[colorlinks,hyperindex]{hyperref}%

\usepackage{changes}

\providecommand{\U}[1]{\protect\rule{.1in}{.1in}}
\hypersetup{
colorlinks,
citecolor=blue,
linkcolor=black,
urlcolor=black,
}

\newcommand{\be}{\begin{equation}}
\newcommand{\ee}{\end{equation}}
\newcommand{\beqn}{\begin{eqnarray}}
\newcommand{\eeqn}{\end{eqnarray}}

\begin{document}
\title{Effects of light-mass fermionic dark matter on the equilibrium and stability of white dwarfs}

\author{G. A. Carvalho}
\email{gacarvalho@utfpr.edu.br}
\affiliation{Departamento de F\'isica, Universidade Tecnol\'ogica Federal do Paran\'a, 85722-332 Medianeira, PR, Brazil}
\affiliation{Programa de P\'os-Gradua\c{c}\~ao em F\'isica e Astronomia, Universidade Tecnol\'ogica Federal do Paran\'a, Jardim das Americas, 82590-300, Curitiba, PR, Brazil}
\author{Jos\'e D. V. Arba\~nil}
\email{jose.arbanil@upn.pe}
\affiliation{Departamento de Ciencias, Universidad Privada del Norte, Avenida el Sol 461 San Juan de Lurigancho, 15434 Lima, Peru}
\affiliation{Facultad de Ciencias F\'isicas, Universidad Nacional Mayor de San Marcos, Avenida Venezuela s/n Cercado de Lima, 15081 Lima, Peru}
\author{Jaziel G. Coelho}
\email{jaziel.coelho@ufes.br}
\affiliation{N\'ucleo de Astrof\'isica e Cosmologia (Cosmo-Ufes) \& Departamento de F\'isica, Universidade Federal do Esp\'irito Santo, 29075-910, Vit\'oria, ES, Brazil}
\affiliation{Divis\~ao de Astrof\'isica, Instituto Nacional de Pesquisas Espaciais, Avenida dos Astronautas 1758, 12227-010, S\~ao Jos\'e dos Campos, SP,
Brazil}


\begin{abstract}

White dwarfs (WDs) can be used as laboratories to test strong gravity and high-density regimes, once their equation of state is not so uncertain as the one of neutron stars. This makes them also a useful tool to constrain dark-matter models. In this work, we study dark matter white dwarfs (DMWD) composed of white dwarf matter admixed with fermionic dark matter in a two-fluid general relativistic framework. Dark matter particles are considered to have masses between $0.1-10$ GeV. The equilibrium configurations and stability are derived, showing that the DMWD can be more compact, with masses around 1.3 $M\odot$ and radii around 500 km. The increasing compactness leads to changes in the fundamental modes of radial oscillations ($\sim20\%$ for 0.1 GeV DM), which produce detectable shifts in GW frequencies. 
The interplay between dark matter and normal matter thus provides a compelling avenue for interpreting deviations in observed WD properties and for placing constraints on DM characteristics through astrophysical observations. 
\end{abstract}
\maketitle

\section{Introduction}\label{sec:int}

The nature of dark matter (DM) particles is largely unknown, despite the numerous observational evidence of its existence, such as rotation curves of galaxies \cite{Rubin1970Feb,Navarro1996May,Catena2010Aug}, large-scale structure formation \cite{Springel2005Jun}, and gravitational lensing \cite{Massey2007Jan}. The DAMA/LIBRA and CoGeNT experimental results have shown consistency in detecting light dark matter particles with a mass of $\sim 10$ GeV \cite{CoGeNTCollaboration2011Mar,Bernabei2008Aug,Bernabei2010May}, but in conflict with results from \cite{CDMSCollaboration2011Mar,XENON100Collaboration2010Sep},  where no detection was found. A new phase of DAMA/LIBRA and new underground programs, such as ANAIS \cite{Bernabei2020Sep} and COSINE \cite{COSINE-100Collaboration2022Sep} are currently running, and novel constraints to the nature of dark matter will be given. 

If one investigates their effects on the structure and properties of compact objects, one can gain more insight into the light nature of DM particles. This kind of analysis can also give restrictions to DM models, such as their mass or self-interactions, parameters that are often studied over a wide interval. For example, fermionic models of dark matter include DM masses from $10^{-5}$ to $10^{5}$ GeV \cite{Deliyergiyev2019Mar,Curtin2021Mar,Bell2021Oct}. 

Compact objects have been used to study the triggering of supernovae from dark matter admixed WDs that are compatible with sub-luminous type Ia supernovae \cite{Leung2015Oct}. Another study raised the possibility of DM-electron interactions leading to DM capture, which in turn deposits energy into hydrogen-atmosphere WDs, leading to variations in the period change rate of pulsation modes \cite{Niu2018Nov}. Asteroseismology could allow us to probe the existence of DM-electron interactions$^{[1]}$\footnotetext[1]{While we neglect DM-electron interactions here, their inclusion (e.g., via a Yukawa coupling) could modify the EoS at low densities. This scenario will be explored in future work, as it may affect pulsation modes~\cite{Niu2018Nov} and heating rates \cite{Horowitz2020Oct}.}. Additionally, authors of \cite{Panotopoulos2020Jun} have used the luminosity of observed WDs in the globular cluster Messier 4 to constrain the independent cross-section of self-annihilating captured sub-GeV DM particles ($\sim 5$ GeV). \cite{Horowitz2020Oct} also investigated the heating from dark matter annihilation to explain the luminosity of some massive WDs, and found that DM capture is unlikely to provide significant heating in massive WD.

The DM admixed WD models are also studied. By numerical integration of the modified Tolman-Oppenheimer-Volkoff equations and radial oscillation equations, the equilibrium configurations and the stability against small radial perturbations are respectively investigated. For instance, in \cite{Ryan2022Jun} DM admixed compact objects were studied using a fermionic degenerate equation of state for both normal and dark matter by considering a scaling relation of fermion masses $m_{\rm NM}/m_{\rm DM}$. In this approach, the rest mass energy of nucleons is neglected, which for WDs contributes more to the energy density than the electron relativistic kinetic energy. A similar scaling between fermionic masses was also used in \cite{Narain2006Sep,Deliyergiyev2019Mar}.

In \cite{Leung2013Jun}, the WDs admixed with DM are also analyzed. In this work, the authors used the Akmal-Pandharipande-Ravenhall equation of state \cite{Akmal1998Sep} for high-density nuclear matter with SLY4 \cite{Douchin2001Dec} and Baym-Pethick-Sutherland \cite{Baym1971Dec} equations of state (EoS) at lower densities. At this point it is worth citing that DM accretion onto compact stars can occur, e.g., due to gravitational capture through elastic scattering \cite{Gould1987Oct,Bertone2008Feb,Kouvaris2010Sep,Sanchis-Gual2022Apr}, adiabatic contraction during gravitational collapse of the progenitor \cite{Brito2015Sep} and accumulation from dense DM environments through DM inelastic scattering on nuclei \cite{McCullough2010Apr,Bramante2015Sep}.

The DM effects on the properties of neutron stars (NSs) were also investigated. For example, the effect of DM particles was also included to study the structure, tidal deformability, and moment of inertia of neutron stars described by relativistic mean-field models with short-range correlations \cite{Lourenco2022Jan,Lourenco2022Aug,Dutra2022Dec}. The results satisfactorily reproduce the constraints on the mass-radius relation derived from analyses of the data from the NICER mission, LIGO collaboration, and mass measurements from radio observations. The effects of DM particles were also extended to hybrid stars in \cite{Lenzi2023Mar}. Admixed NSs including different models for the DM component have also been studied in detail in several papers \cite{Leung2011Nov,Kain2021,Leung2022Jun,Mariani2024Jan,Thakur2024Dec,Grippa2025Feb,Mahapatra2025May,Thakur2025May}. For instance, Leung et. al \cite{Leung2011Nov,Leung2022Jun} investigate the two-fluid formalism in the context of bosonic and fermionic dark matter, demonstrating that even small DM fractions can significantly alter the mass-radius relation of NSs. More recently, Kain \cite{Kain2021} studied DM admixed NSs for two choices of dark matter: a free Fermi gas and mirror dark matter and analyzed their dynamical stability, while Mariani et al. \cite{Mariani2024Jan} studied the equilibrium configurations of dark matter-admixed stars under self-interacting dark matter scenarios. The very recent comprehensive review by Grippa et al. \cite{Grippa2025Feb} consolidates many of these developments, highlighting the theoretical frameworks and observational consequences associated with dark matter inside compact stars.

Unlike previous studies \cite{Ryan2022Jun,Leung2013Jun} that neglected nucleon rest-mass contributions or focused on heavier DM particles, this work (1) includes the full energy density from ions, (2) examines light fermionic DM (0.1-10 GeV), and (3) simultaneously analyzes both equilibrium configurations and dynamical stability through radial oscillations - crucial for identifying observable signatures in compact DM-admixed WDs. Also, this work aims to fill the gap by systematically studying dark matter effects in the structure and stability of WDs, particularly in the ultramassive regime. Such configurations have also been proposed to arise from double white dwarf (DWD) mergers, potentially explaining recently observed isolated, massive, fast-rotating, and highly magnetized WDs \cite{Sousa2022, Sousa2023}.

This work is structured as follows: in Section \ref{sec2}, we present the hydrostatic equilibrium equations, the EoS for both normal matter (with explicit inclusion of nucleon rest-mass energy) and fermionic dark matter, and the formalism for radial stability analysis. In Section \ref{sec3}, the results are discussed. In Section~\ref{sec4}, we translate these findings into observable signatures, including surface gravity anomalies ($\Delta\log g \approx 0.2$ dex), gravitational redshifts ($z_{\text{surf}} \approx 0.1$), and GW frequency modifications detectable by next-generation instruments. We conclude in Section~\ref{sec5} by discussing implications for sub-Chandrasekhar WD populations and propose future directions, including DMWD merger simulations and coupling to asteroseismology data.

\section{The formalism}\label{sec2}

\subsection{The hydrostatic equilibrium equations}

In this work, we assume that the fluid contained in the star is composed of two fluids that interact only gravitationally. The line element that describes the spherically symmetric object can be written as $ds^2={\rm e}^\nu dt^2 - {\rm e}^\lambda dr^2 - r^2 d\theta^2 - r^2\sin^2\theta d\phi^2$. Through these considerations, the Tolman-Oppenheimer-Volkoff hydrostatic equilibrium equations can be placed as \cite{Tolos2015Dec,Mukhopadhyay2016Apr},
\begin{eqnarray}\label{eq1}
&&\frac{dp_{\rm NM}}{dr}= -\frac{GM\epsilon_{\rm NM}}{r^2}\left(1+\frac{p_{\rm NM}}{\epsilon_{\rm NM}} \right) \nonumber \\
&&\left(1+4\pi r^3\frac{(p_{\rm NM}+p_{\rm DM})}{M}\right) \left(1-\frac{2M}{r}\right)^{-1},\\
&&\frac{dp_{\rm DM}}{dr}= -\frac{GM\epsilon_{\rm DM}}{r^2}\left(1+\frac{p_{\rm DM}}{\epsilon_{\rm DM}} \right) \nonumber \\
&& \left(1+4\pi r^3\frac{(p_{\rm NM}+p_{\rm DM})}{M}\right) \left(1-\frac{2M}{r}\right)^{-1},\\
&&\frac{dm_{\rm NM}}{dr}=4\pi r^2 \epsilon_{\rm NM},\\
&&\frac{dm_{\rm DM}}{dr}=4\pi r^2 \epsilon_{\rm DM},\label{eq4}
\end{eqnarray}
with $M=m_{\rm NM}+m_{\rm DM}$ being the sum of mass components, and the sub-indices DM and NM represent dark matter and normal matter, respectively. To obtain hydrostatic equilibrium solutions, we must define the equations of state for normal and dark matter components, which are detailed in the next section. The set of equations \eqref{eq1}$-$\eqref{eq4} are integrated from the center ($r=0$) towards the star's surface ($r=R$). In $r=0$, we use $m_{\rm NM}(r=0)=m_{\rm DM}(r=0)=0$, $p_{\rm NM}(r=0)=p_{\rm NM}^C$, and $p_{\rm DM}(r=0)=p_{\rm DM}^C$. The surface of the star $r=R$ is found when both pressures vanish, i.e., $p_{\rm NM}(R)=p_{\rm DM}(R)=0$.

\subsection{The equations of state}

\subsubsection{Normal matter}

The relativistic degenerate electron gas model can be applied to study WD matter. In this model, the energy density is essentially given by the contribution of the rest mass energy density of the ions and the kinetic relativistic electron energy density. In contrast, the electron degeneracy pressure gives pressure. This is called the Chandrasekhar model and can be improved to include lattice effects and self-interaction corrections. An important point that has sometimes been dismissed in previous works (see, e.g., \cite{Narain2006Sep,Leung2013Jun,Deliyergiyev2019Mar,Ryan2022Jun,Barbat2024Jul}) on DMWD is the contribution to the energy density from the rest mass of the ions. Another unaddressed point is that the energy density is limited by inverse beta decay and nuclear processes that occur at high densities \cite{Hamada1961Nov,Chamel2013Oct,Chamel2015Jul,Otoniel2019Jul}. In this work, the pressure and total energy density of the normal matter are given, respectively, by
\begin{eqnarray}
    p(k_F)&=& \frac{1}{3\pi^2\hbar^3} \int_0^{k_F} \frac{k^4}{\sqrt{k^2+m_e^2}}dk, \\
    \epsilon(k_F)&=& \epsilon_i +\epsilon_e, \\ 
    &=& \frac{m_N\mu_e k_F^3}{3\pi^2\hbar^3} + \frac{1}{\pi^2\hbar^3}\int_0^{k_F}\sqrt{k^2+m_e^2}k^2dk, 
\end{eqnarray}
where $m_e$ is the electron mass, $\hbar$ is the reduced Plank mass, $k_F$ is the Fermi momentum, $\mu_e$ is the ratio between mass and atomic numbers for ions and $m_N$ is the nucleon mass.

\begin{figure}
    \centering
    \includegraphics[width=1.0\linewidth]{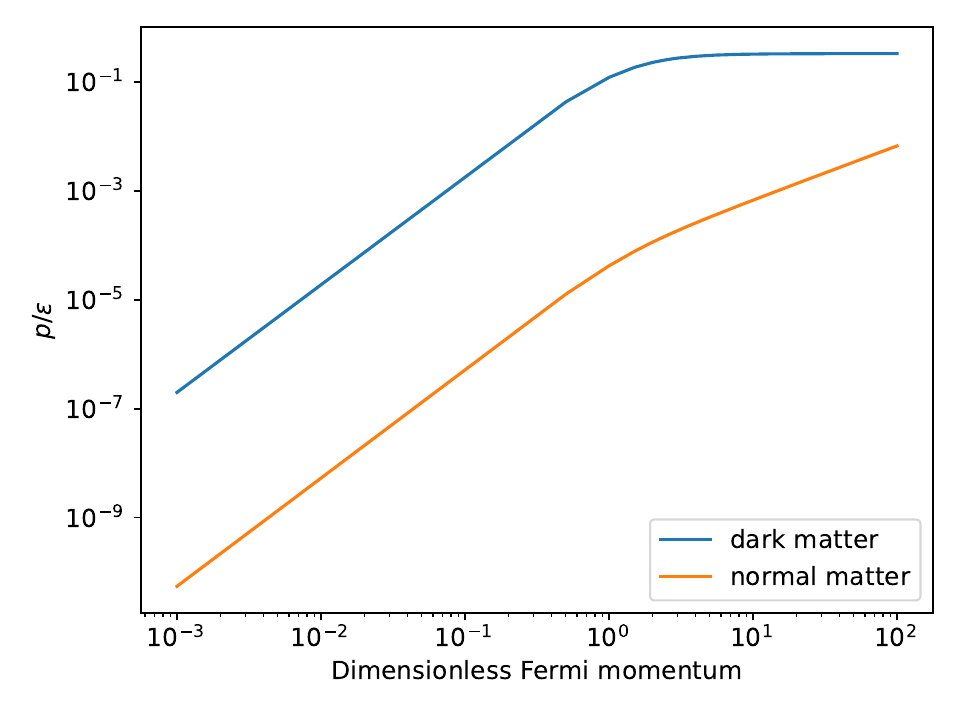}
    \caption{Pressure and energy density ratio as a function of the dimensionless Fermi momentum for both normal and dark matter equations of state.}
    \label{fig:eos}
\end{figure}

\begin{figure*}
  \includegraphics[scale=0.5]{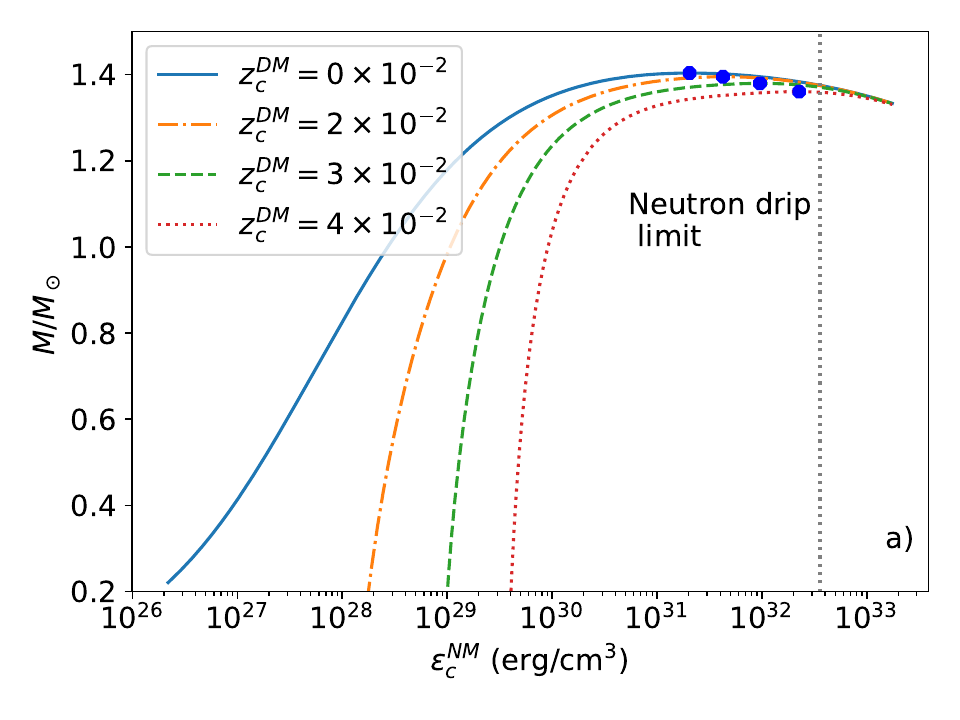}
  \includegraphics[scale=0.5]{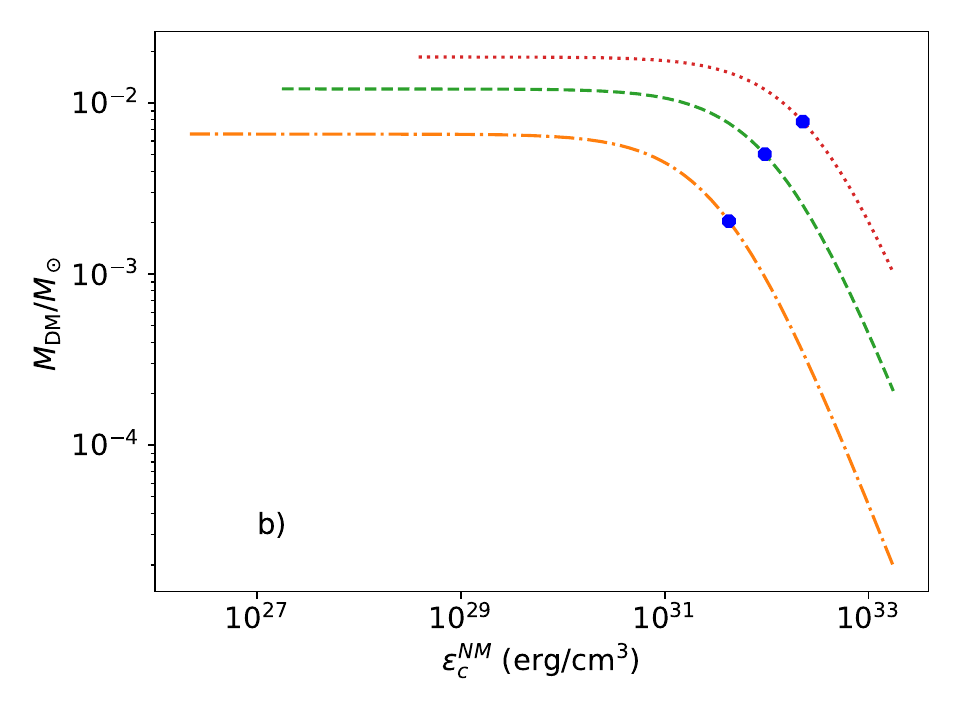}
  \includegraphics[scale=0.5]{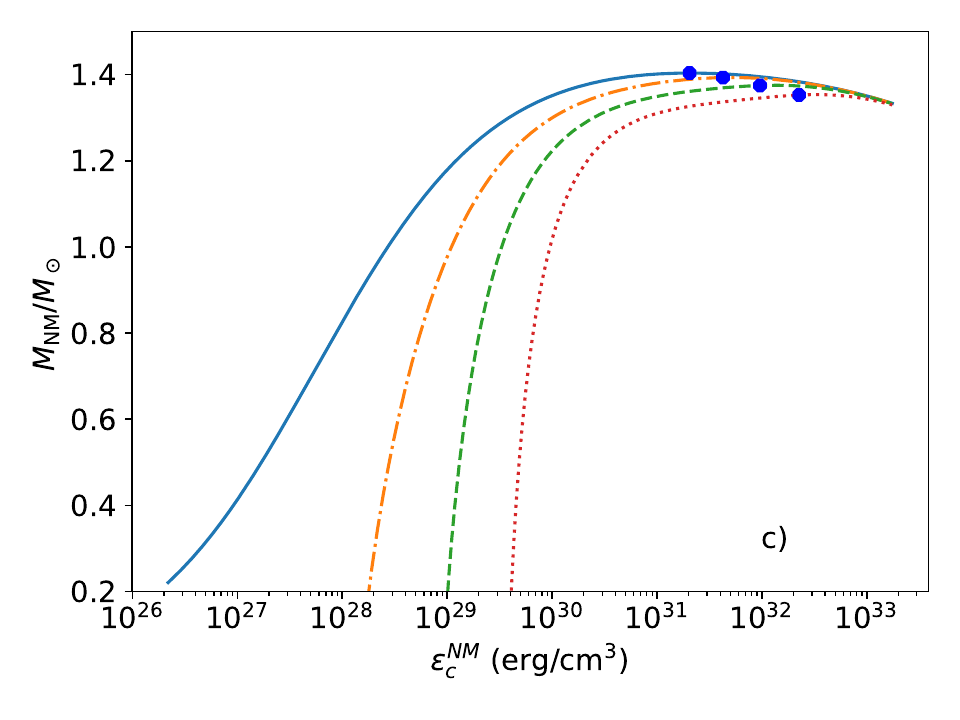}
  \includegraphics[scale=0.52]{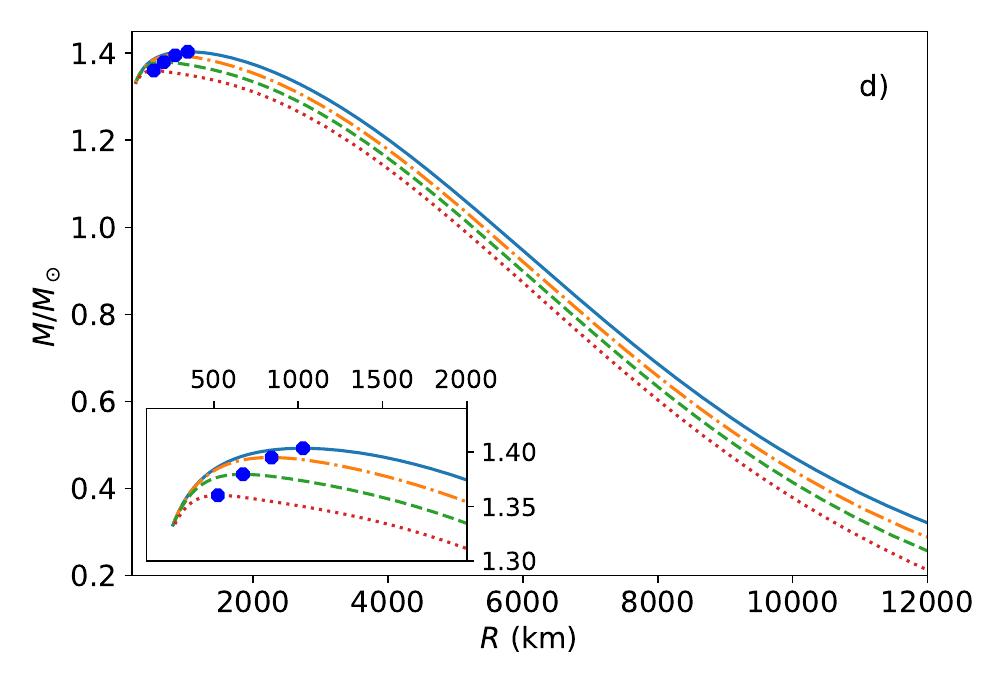}
  \includegraphics[scale=0.52]{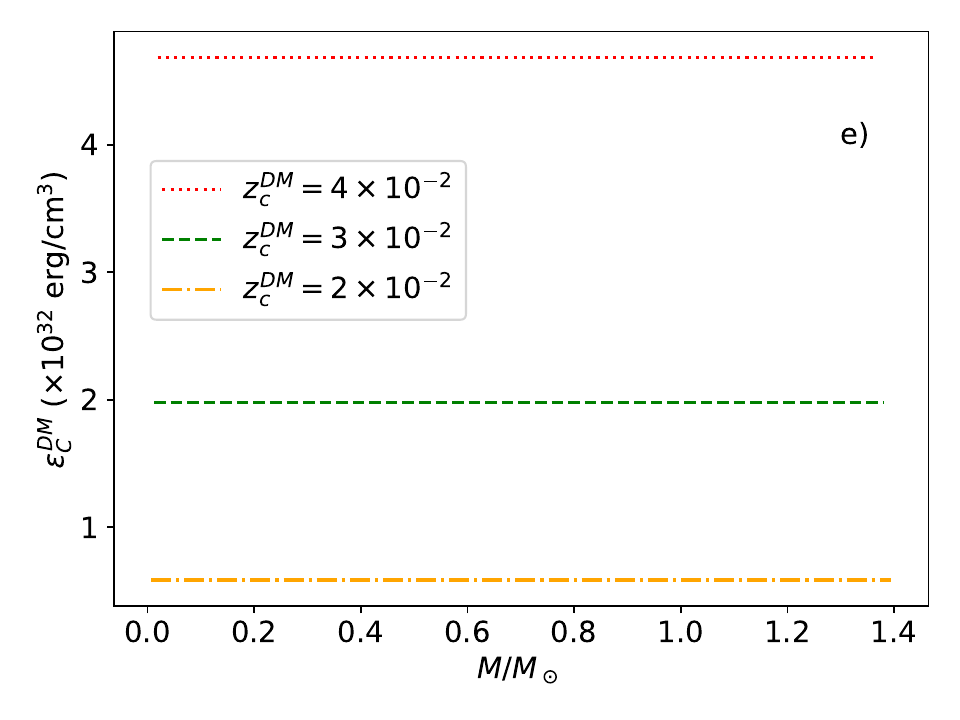}
  \caption{\label{mr1} Panel a): Total mass versus normal matter central density. Panel. b): Dark matter component of the star's mass as a function of normal matter central density. Panel c): Normal matter component of the mass of the star as a function of normal matter central density. Panel d): Total mass versus total radius of the DMWD stars. Panel e): Dark matter central energy density versus total mass. In all panels, we consider $m_{\rm DM}=1$ GeV and different values of central Fermi momentum for the dark matter content.}
\end{figure*}

\subsubsection{Dark matter}

The pressure and energy density of the dark matter fluid are described by a relativistic fermionic model at zero temperature; which corresponds to a cold dark matter model. The maximum temperature inside white dwarfs is in the scale of $10^9$ K, corresponding to an energy scale of $100$ keV, while we consider dark particle masses between $0.1-1$ GeV, which shows the feasibility of the zero temperature approximation. The dark matter pressure and energy density are given, respectively, by
\begin{eqnarray}
p(k_F)&=& \frac{1}{3\pi^2\hbar^3} \int_0^{k_F} \frac{k^4}{\sqrt{k^2+m_f^2}}dk, \\
\epsilon(k_F)&=& \frac{1}{\pi^2\hbar^3}\int_0^{k_F}\sqrt{k^2+m_f^2}k^2dk, 
\end{eqnarray}
where $k_F$ represents the Fermi momentum of DM particles and $m_f$ depicts the dark matter particle mass.

Defining a dimensionless Fermi momentum for dark matter particles as $z=k_F/m_f$, the pressure and energy density can be written in the following form
\begin{eqnarray}
    p(k_F)&=& \frac{m_f}{24\pi^2\hbar^3} \left[(2z^3-3z)\sqrt{z^2+1}+3\sinh ^{-1}(z)\right], \nonumber\\
    \epsilon(k_F)&=& \frac{m_f}{8\pi^2\hbar^3} \left[(2z^3+z)\sqrt{z^2+1}-\sinh ^{-1}(z)\right]. 
\end{eqnarray}

\begin{figure*}
  \includegraphics[scale=0.5]{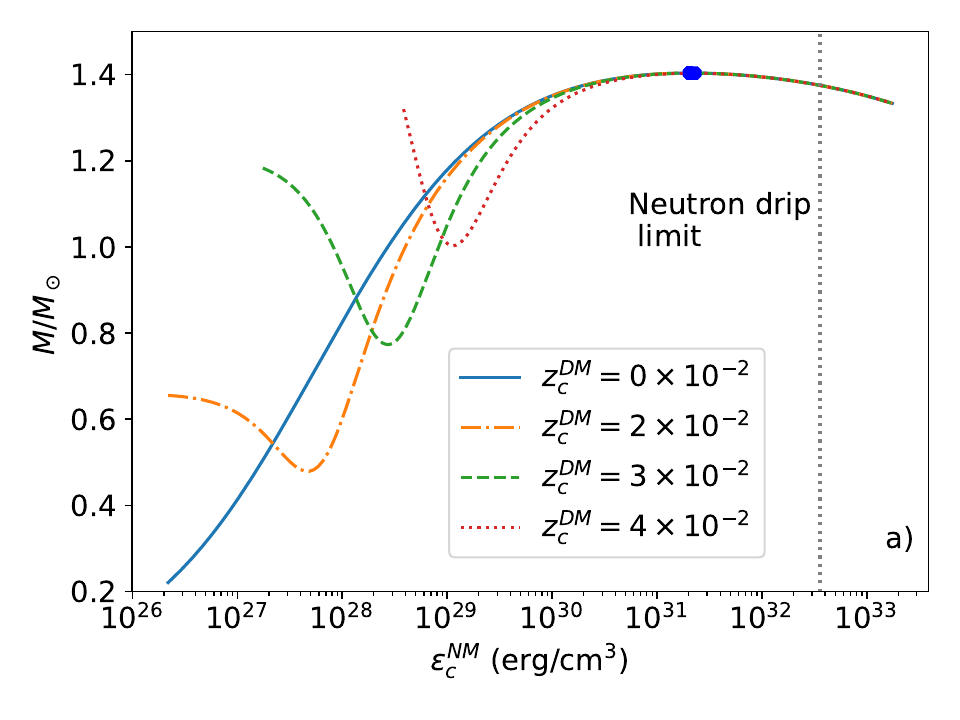}
  \includegraphics[scale=0.5]{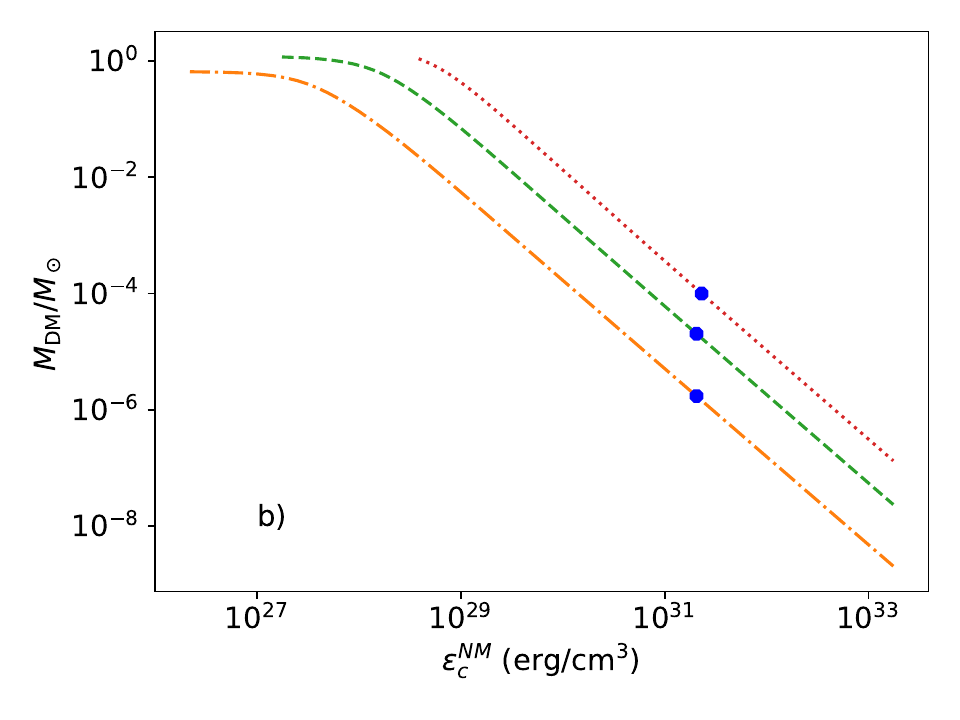}
  \includegraphics[scale=0.5]{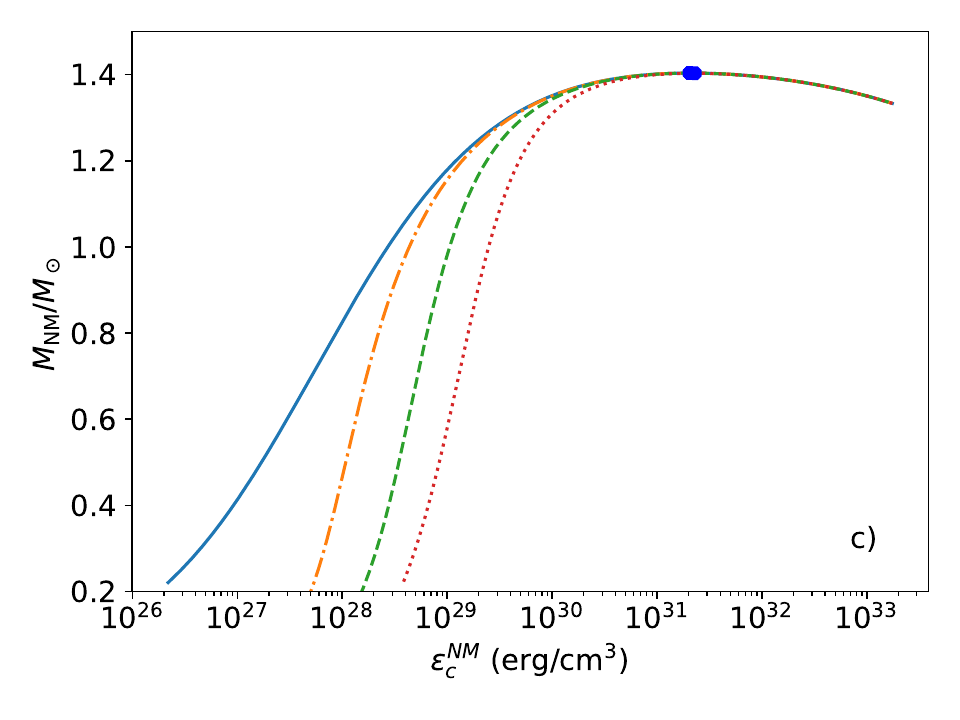}
  \includegraphics[scale=0.5]{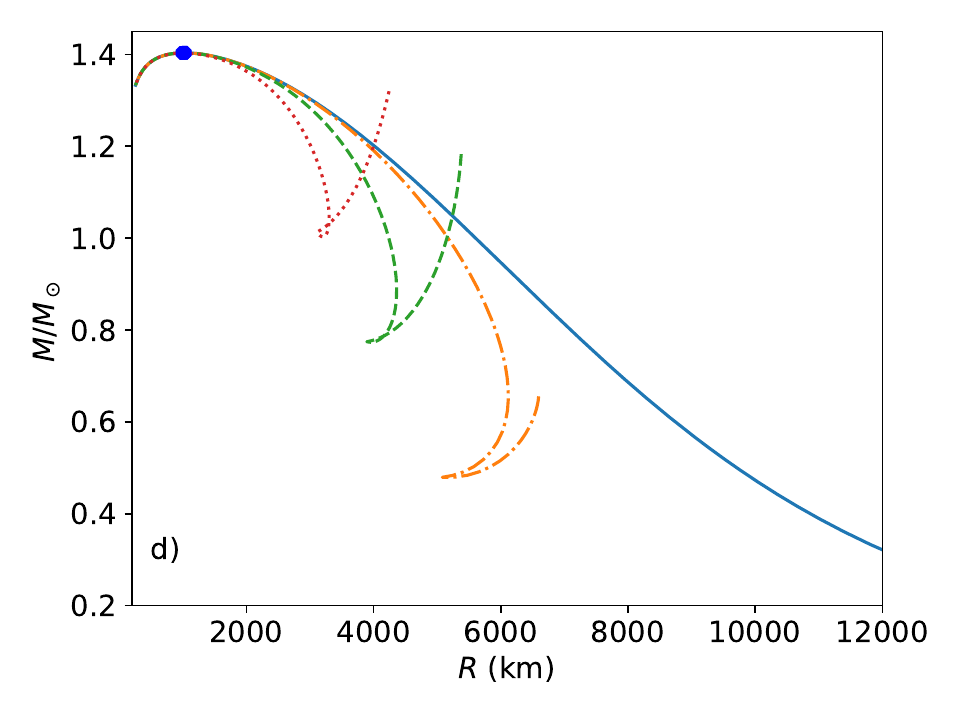}
  \includegraphics[scale=0.52]{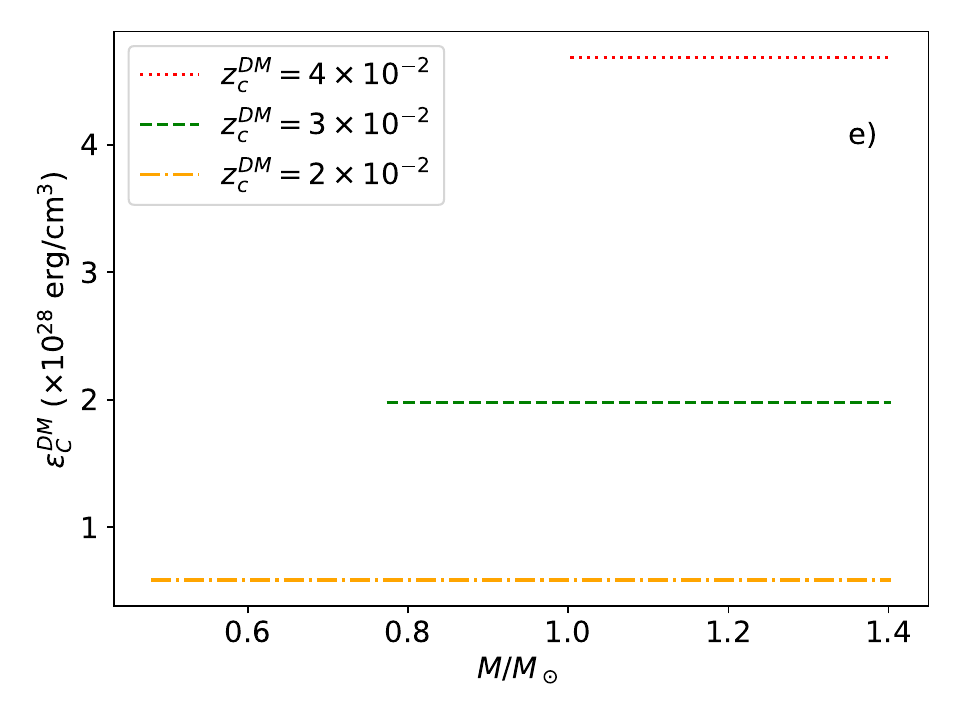}
  \caption{\label{mr01} Same as Fig. \ref{mr1}, but for $m_{\rm DM}=0.1$ GeV.}
\end{figure*}

\subsection{Radial Oscillation Equations}

The radial stability equations are derived by perturbing space-time metric and fluid variables, while preserving the background spherical symmetry for the object and maintaining only first-order terms \cite{Chandrasekhar1964}. These equations can be placed into two first-order differential equations, one for the relative radial displacement $\xi$ and another for the Lagrangian perturbation $\Delta p_r$, which defines a Sturm-Liouville problem for the determination of radial oscillation frequencies, $\omega^2$. With the help of a trial function for the relative radial displacement, $\xi=re^{\nu/2}$, one can then calculate the square of oscillation frequencies as 
\begin{equation}\label{ro1}
    \omega^2= \frac{\mathcal{Z}}{\mathcal{D}},
\end{equation}
where $\mathcal{Z}$ and $\mathcal{D}$ are given by
\begin{eqnarray}
\mathcal{Z}&=& 4 \int_0^R e^{(3\nu+\lambda) / 2} r^3 \frac{d p}{d r} d r \nonumber\\  &+& 9\int_0^R e^{(3\nu+\lambda) / 2} r^2 \Gamma p d r \nonumber \\
&-&\int_0^R \frac{e^{(3\nu+\lambda) / 2} r^4}{(p+\epsilon)}\left[\frac{d p}{d r}\right]^2 d r  \nonumber \\
&+& 2\pi \int_0^R e^{(3\nu+3\lambda)/2} p(p+\epsilon) r^4 d r, \label{ro2} \\
\mathcal{D}&=&\int_0^R e^{(3\lambda + \nu) / 2}(p+\epsilon) r^4 d r,\label{ro3}
\end{eqnarray}
where $\lambda$ and $\nu$ are the metric potentials, $p=p_{\rm NM}+p_{\rm DM}$ and $\epsilon=\epsilon_{\rm NM}+\epsilon_{\rm DM}$. It is worth citing that the metric potentials and the $\Gamma$ can be obtained by calculating, respectively 
\begin{eqnarray}
    \frac{d\nu}{dr} &=& -\frac{2}{(p+\epsilon)}\frac{dp}{dr},\\
    {\rm e}^\lambda &=& \left(1-\frac{2M}{r}\right)^{-1}\\
    \Gamma&=& \frac{\epsilon}{p}\frac{dp}{d\epsilon}.
\end{eqnarray}
Hence, after obtaining the functions of the radial pulsation equations on each hydrostatic equilibrium configuration, the radial pulsation fundamental modes can be derived from Eqs. \eqref{ro1}$-$\eqref{ro3}.

\begin{figure*}
  \includegraphics[scale=0.5]{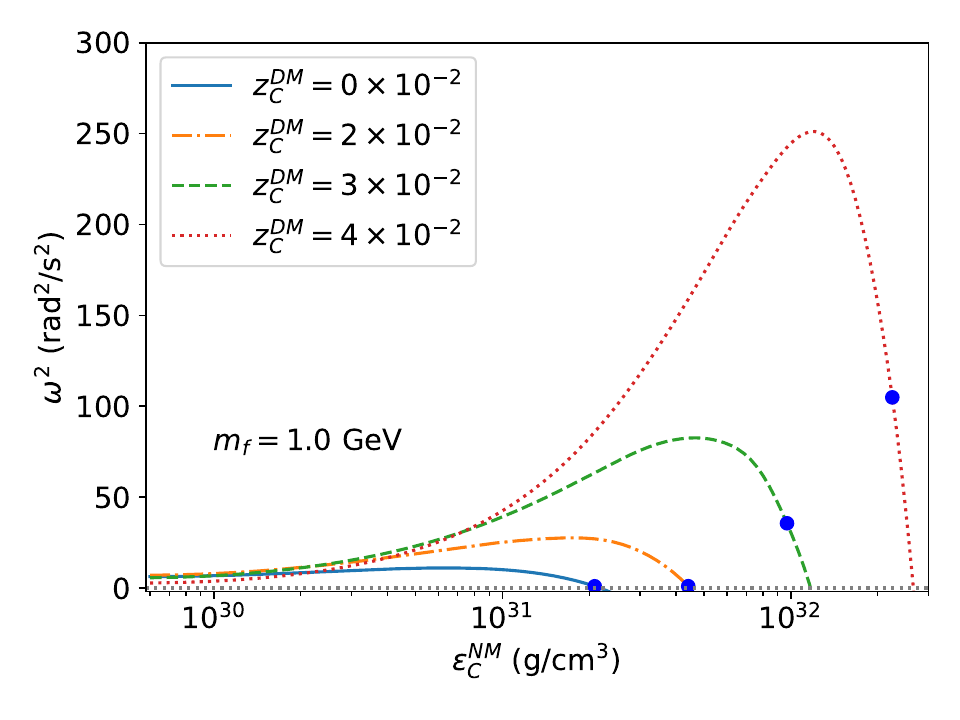}
  \includegraphics[scale=0.5]{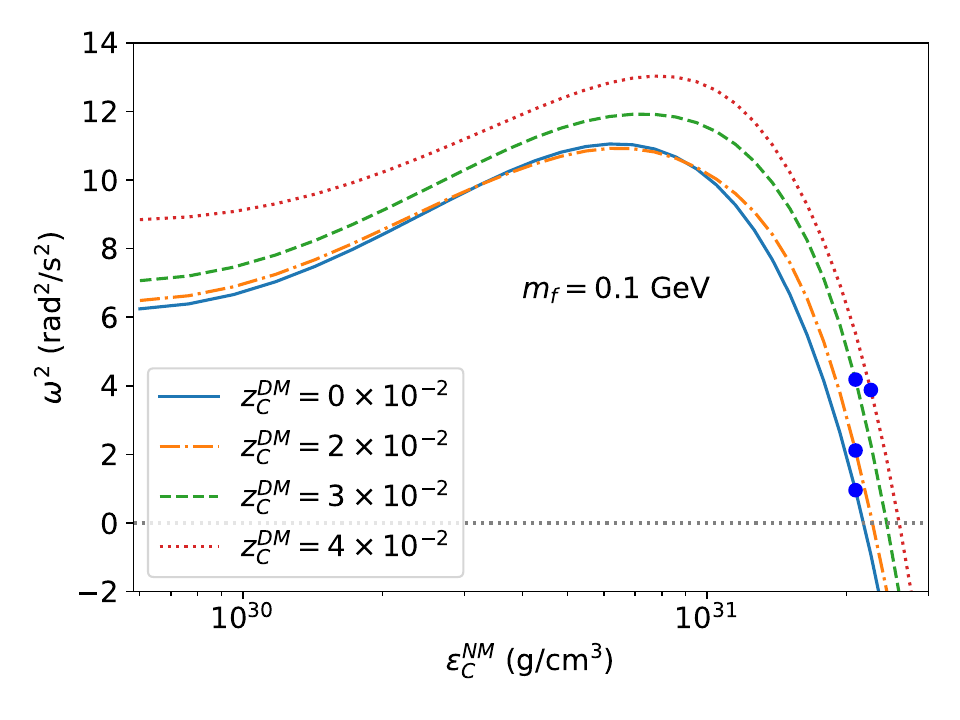}
  \caption{\label{radial} Radial oscillation frequencies as a function of the normal matter central density. The curves correspond to different values of central Fermi momentum for the dark matter content. Two values of dark matter particles are considered as displayed in each graph.}
\end{figure*}

\section{Results}\label{sec3}

In Figure \ref{fig:eos}, the ratio between pressure and energy density is plotted as a function of the dimensionless Fermi momentum for both the standard and dark matter equations of state. It is evident that this ratio is generally larger for the dark matter EoS. This behavior arises because the energy density of normal matter includes the rest mass energy of ions, significantly increasing its overall scale. As a result, the central dimensionless Fermi momentum values used for dark matter in this work are of the order of $10^{-2}$, whereas for normal matter they span a broader range, from $10^{-1}$ to $10^2$. The chosen scale of $10^{-2}$ for the central Fermi momentum of the dark matter component reflects a balance between physical relevance and stability: smaller values yield negligible impact on the mass-radius relation, while significantly larger values lead to gravitational collapse, producing stars with extremely small radii (on the order of $\sim 1$ m). Naturally, this characteristic scale depends on the assumed dark matter particle mass, and would shift accordingly under different mass assumptions.

Since we are plotting the ratio $p/\epsilon$, the specific fermion masses cancel out and do not affect the shape of the curves. Nonetheless, it is important to note that we adopt dark matter particle masses in the range of 0.1$-$1 GeV, while the electron mass is approximately $0.5$ MeV. This large mass disparity further illustrates the scale difference between the two equations of state. Our choice of fermionic DM masses ($0.1-1$ GeV) aligns with the parameter space least constrained by direct detection experiments. For instance, XENONnT~\cite{2023PhRvL.131d1003A} excludes spin-independent DM-nucleon cross-sections above $10^{-45} \rm ~cm^2$ for $m_f\sim 1 \rm ~GeV$, but there are numerous experiments aiming to probe sub-GeV dark matter. Astrophysical probes like WD heating \cite{Horowitz2020Oct} and pulsation timing~\cite{Niu2018Nov} thus offer complementary constraints.

In panel (a) of Figure \ref{mr1}, we display the total mass of the white dwarf as a function of the central energy density of the normal matter. Each curve corresponds to a different central Fermi momentum value for the dark matter component. The vertical dotted line marks the neutron drip density threshold at approximately $3.6\times 10^{32}$ erg/cm$^{3}$. Panels (b) and (c) show the contributions of the normal and dark matter, respectively, to the total mass of WDs. Panel (d) shows the total mass as a function of stellar radius{\bf , while panel (e) shows the dark matter central energy density as a function of total mass}. Maximum mass configurations are indicated by blue circles across all panels. In this figure, the dark matter particle mass is set to $1$ GeV.

From Figure \ref{mr1}, it can be seen that increasing the central Fermi momentum of the dark matter leads to a greater dark matter mass contribution inside the WD, while simultaneously reducing the normal matter contribution. Because the total mass is primarily dominated by the normal matter component, the overall mass decreases as the DM content increases. Notably, larger dark matter Fermi momenta result in more compact configurations, i.e., lower maximum masses and smaller radii, though they require higher central energy densities in the normal matter component. {As the central Fermi momentum for the dark component is fixed constant the central energy density is also constant in each case.}

Taking the neutron drip threshold into account, the smallest maximum total mass found is $1.35 M_\odot$, corresponding to a radius of approximately $455$ km. This indicates that dark matter admixture alters the structure of WDs by increasing their compactness, albeit with slightly reduced masses. This structural change is primarily due to the additional pressure contributed by DM, which allows the star to support more mass against gravity, despite its smaller radius.

It is worth noting that the presence of DM within WDs results in sub-Chandrasekhar limiting masses. However, even in these cases, the maximum mass does not fall below $1.35 M_\odot$, which represents only a $3.7\%$ reduction compared to the maximum mass of a standard WD without dark matter ($1.41 M_\odot$). Therefore, equilibrium configurations alone cannot explain the existence of sub-Chandrasekhar white dwarfs with masses below $1.3 M_\odot$.

In Figure \ref{mr01}, we present mass$-$radius and mass-central energy density relations similar to those in Figure \ref{mr1}, but for a dark matter particle mass of $0.1$ GeV. In this case, DM significantly affects the WD structure at lower central energy densities. At higher densities, near the maximum mass configurations, the curves nearly overlap with the no-dark-matter case, indicating negligible structural impact in that regime.

Conversely, in the low-density regime, dark matter begins to dominate the total mass of the star, leading to significant deviations in the mass$-$radius and mass$-$central energy density relations. For instance, with a central dimensionless Fermi momentum of $z_C^{DM}=4\times 10^{-2}$, the minimum total mass of a dark matter-admixed WD is around $1 M_\odot$. This implies that, under such conditions, white dwarfs with masses below $1 M_\odot$ would not be possible if they contained this level of DM.

While $\omega^2 > 0$ confirms stability for the fundamental mode, see Figure \ref{radial}, higher-order modes ($n \geq 1$) could reveal DM-induced instabilities at densities above neutron drip. 
However, a comprehensive analysis of eigenfunctions and boundary conditions is needed for $n \gg 1$. The {3.7\%} mass reduction at the stability limit (Fig.~\ref{mr1}a) implies DMWDs may avoid supernova ignition until reaching higher densities than standard WDs, consistent with subluminous Type~Ia scenarios \cite{Leung2015Oct}.

 \section{Astrophysical implications}\label{sec4}
 
One of the key astrophysical implications of these results is that, as stars become more compact, both the redshift and surface gravity are altered in dark matter-admixed white dwarfs (DMWDs). These two quantities - redshift and surface gravity - are sometimes measurable through observational instruments. Notably, the Gaia catalog of nearby stars includes around $60$ apparently faint blue-white dwarfs, some of which have been confirmed by \cite{Scholz2022Feb}. These objects offer a promising opportunity to study ultra-massive WDs, as their position in the color-magnitude diagram deviates from theoretical models of such stars \cite{Althaus2022Dec}. The unusual properties of these faint blue-white dwarfs may be explained by increased compactness resulting from the presence of DM.

For a DMWD with $M = 1.3\,M_{\odot}$ and $R = 455$~km (Fig.~\ref{mr1}d), surface gravity increases by $\Delta \log g \approx 0.2$~dex compared to a Chandrasekhar WD of the same mass. This aligns with the anomalous $\log g$ values of Gaia's faint blue WDs with surface gravities $\log g \approx 9.0$--$9.5$ for masses $>1.2\,M_{\odot}$, significantly higher than predictions from standard cooling models ($\log g \approx 8.7$--$9.0$). Similarly, the gravitational redshift $z_{\mathrm{surf}} \approx 0.1$ would be detectable via shifted absorption lines in UV spectra (e.g., {\it HST}/COS). Similar anomalies are observed in ultramassive WDs \cite{Althaus2022Dec}.

Furthermore, as shown in Figure \ref{radial}, the stability of these stars under radial perturbations is demonstrated by the fact that the maximum mass configurations (indicated by blue circles) occur prior to the point where $\omega^2 < 0$, thus obtaining a new arm of stable stars beyond the maximum mass point. This is a result similar to those found for dark matter-admixed NSs \cite{Kain2021,Caballero2024} and other compact stars, for instance, in both general relativity \cite{Pereira2018,Lugones2023JCAP} and in extended theories of gravity \cite{Arbanil2015,Arbanil2016JCAP}. However, in WDs, the eigenfrequency of oscillation is approximately a thousand times smaller than that found for NSs and other compact objects. In that sense, since the eigenfrequency of oscillation near the maximum mass point remains close to $\omega = 0$, we may infer that the Chandrasekhar stability criterion, based on the conditions $dM/d\rho_c > 0$ for stability and $dM/d\rho_c < 0$ for instability, can still be applied to equilibrium sequences of WDs admixed with DM. In addition, we considered two values for the dark matter particle mass, as specified in the figure. The results show that when the dark matter particle mass is $0.1$ GeV, the variations in the oscillation frequencies are significantly larger compared to the $1$ GeV case. In \cite{Tang2023May}, it was found that crystalline cores within WDs do not alter the fundamental modes of radial oscillations, but instead affect the tidal deformability. In contrast, our findings reveal that the fundamental modes are sensitive to both the presence and mass of DM within WDs, suggesting that the tidal deformability of DMWDs may differ substantially from that of ordinary white dwarfs. These dark matter effects may leave distinct signatures in the gravitational wave emissions of such stars.

It's worth mentioning that the fundamental mode frequency $\omega_0$ for DMWDs (Fig.~\ref{radial}) shifts by up to $20\%$ for $m_f = 0.1$~GeV, potentially detectable by \textit{LISA} for nearby binaries. Tidal deformability $\Lambda$ (not computed here) is expected to decrease due to compactness, altering GW phasing in mergers. If DMWDs have $\Lambda \propto R^5$ like NSs, their mergers would show phase shifts in GWs~\citep[see][]{2025JHEAp..45..333N}. 

These DM effects may leave distinct signatures in the GW emissions of such stars and in their thermal and rotational evolution. Recent studies have shown that DWD mergers can produce massive, magnetic WDs like J2211+1136 \citep{Sousa2022}. Interestingly, even in the absence of DM, optical transients arising from double WD mergers leading to ultramassive remnants have been theoretically modeled and are expected to be detectable in forthcoming surveys such as LSST of Vera Rubin \cite{Sousa2023}.

Our results for DMWDs with masses of approximately $1.3\,M_\odot$ and radii of about $500$ km are consistent with the properties of white dwarfs formed in  DWD mergers, such as J2211+1136 and ZTF J1901+1458 \cite{Sousa2022}. The compactification induced by DM could account for deviations in the observed values of $\log g$ and gravitational redshift in such ultramassive WDs. Since DWD mergers produce dynamical ejecta with masses on the order of $10^{-3}\,M_\odot$ \cite{Sousa2023}, the presence of DM in the central remnant may modulate the accretion rate and the energy injection into the ejecta, consequently altering the luminosity and duration of the associated optical transient. Future simulations should explore how DM affects the rotational evolution of DWD merger remnants and their electromagnetic signatures, potentially combining data from the LSST and GWs from LISA.

The presence of fermionic DM in WDs introduces additional energy transport mechanisms that could alter their thermal evolution. For light DM ($m_f \sim 0.1$ GeV), if we assume a DM-electron scattering cross-section $\sigma_{e\chi} \sim 10^{-37}$ cm$^2$ (consistent with sub-GeV constraints from the XENONnT experiment \cite{XENONnT2023}) and a central DM density $\rho_{\rm DM} \sim 10^8$--$10^9$ g/cm$^3$ (from our Fig.~2), the additional heat conduction might delay cooling by $\sim 1$--$10\%$ during the first Gyr. This estimate follows from scaling arguments similar to \cite{Horowitz2020}, though the exact effect depends strongly on the uncertain velocity dependence of $\sigma_{e\chi}$.
If DM annihilates ($\chi\bar{\chi} \to e^+e^-$), adopting $\langle \sigma v \rangle \sim 10^{-26}$ cm$^3$/s \cite{Steigman2012} and our maximum $\rho_{\rm DM}$ values, the heating luminosity would be $L_{\rm DM} \sim 10^{-3} L_\odot$ -- comparable to late-stage WD photon luminosity. While suggestive, we emphasize that this assumes full thermalization of annihilation products, unlike the more conservative bounds in \cite{Leane2021}.
For $m_f = 1$ GeV, both effects become negligible ($\lesssim 1\%$) due to the $\rho_{\rm DM}/m_f$ scaling. Future work should couple our equilibrium models with thermal evolution codes to refine these estimates. Current surveys (SDSS-V, Gaia DR3) and upcoming LSST observations will critically test these predictions through population-level analyses and precision asteroseismology of massive pulsators, potentially yielding the first astrophysical constraints on sub-GeV DM interactions in degenerate matter environments.

\section{Conclusion}\label{sec5}

Our work demonstrates that the presence of dark matter inside white dwarfs significantly influences their structural properties, particularly in terms of mass, radius, and compactness, which are intrinsically related to surface gravity, redshift, and effective temperature. The key effects are strongly dependent on the dark matter particle mass and the central Fermi momentum.

For heavier dark matter particles (1 GeV), the total masses of the stars decrease with increasing dark matter content, whereas the stars become more compact, which requires higher central energy densities in the normal matter component. 

In contrast, for lighter DM particles (0.1 GeV), the impact of dark matter is negligible at high densities but becomes dominant in the low-density regime, setting a minimum mass threshold near 1 $M_\odot$ in some cases. These findings reveal that dark matter-admixed white dwarfs may exhibit sub-Chandrasekhar mass configurations, but equilibrium models alone cannot account for white dwarfs with masses significantly below 1.3 $M_\odot$, as, for example, in the scale of 1 $M_\odot$. 

Our results demonstrate that fermionic DM admixed WDs exhibit distinct mass-radius relations, oscillation spectra, and surface properties, offering testable signatures for multi-messenger astronomy. Future directions include: (i) coupling DM-electron interactions to asteroseismology data, (ii) simulating GW waveforms from DMWD binaries, and (iii) probing DM capture in WDs via infrared excess.

\section*{Acknowledgments}
{\bf We thank the anonymous referee for their valuable suggestions and comments, which have helped improve this work.}
G.A.C. would like to thank CNPq for financial support under process \#314121/2023-4 and Funda\c{c}\~{a}o Arauc\'{a}ria for financial support under NAPI ``Fen\^{o}menos extremos no Universo''.  J.G.C. is grateful for the support of FAPES (1020/2022, 1081/2022, 976/2022, 332/2023) and CNPq (311758/2021-5).

\bibliography{library}

\end{document}